\documentclass[prl,aps,showpacs,twocolumn,floatfix]{revtex4}
\usepackage{epsfig} \usepackage{graphics} \usepackage{bm}
\usepackage{amssymb}
\usepackage{graphicx}
\begin{document}

\preprint{Lebed-PRL}

\title{Non Fermi Liquid Crossovers in a Quasi-One-Dimensional
Conductor in a Tilted Magnetic Field}

\author{A.G. Lebed$^*$}

\affiliation{Department of Physics, University of Arizona, 1118 E.
4-th Street, Tucson, AZ 85721, USA}

\begin{abstract}
We consider a theoretical problem of electron-electron scattering
time in a quasi-one-dimensional (Q1D) conductor in a magnetic
field, perpendicular to its conducting axis. We show that inverse
electron-electron scattering time becomes of the order of
characteristic electron energy, $1/\tau \sim \epsilon \sim T$, in
a high magnetic field, directed far from the main crystallographic
axes, which indicates breakdown of the Fermi liquid theory. In a
magnetic field, directed close to one of the main crystallographic
axis, inverse electron-electron scattering time becomes much
smaller than characteristic electron energy and, thus,
applicability of Fermi liquid theory restores. We suggest that
there exist crossovers (or phase transitions) between Fermi liquid
and some non Fermi liquid states in a strong enough tilted
magnetic field. Application of our results to the Q1D conductor
(Per)$_2$Au(mnt)$_2$ shows that it has to be possible to observe
the above mentioned phenomenon in feasibly high magnetic fields of
the order of $H \geq H^* \simeq 25 \ T$.
\end{abstract}


\maketitle

High magnetic field properties of quasi-one-dimensional (Q1D) and
quasi-two-dimensional (Q2D) conductors have been intensively
studied since the discovery of the so-called Field-Induced
Spin-Density-Wave (FISDW) cascades of phase transitions in the Q1D
materials (TMTSF)$_2$X (X=ClO$_4$, PF$_6$, etc.) [1-3]. It is
important that successful theoretical explanations of the FISDW
phases [3-8] were not done in the framework of the traditional
theory of metals but required a novel notion - the so-called
quasi-classical $3D \rightarrow 2D$ dimensional crossover. Later,
different types of quasi-classical $3D \rightarrow 1D \rightarrow
2D$ dimensional crossovers were applied for explanations of such
unusual properties of a metallic phase in Q1D conductors as
Lebed's magic angles and Lee-Naughton-Lebed's oscillations [9].
Note, that general feature of the above mentioned dimensional
crossovers is that electron spectrum changes its dimensionality in
moderate magnetic fields, where the typical "sizes" of electron
trajectories are bigger than the inter-plane distances in layered Q1D
conductors. Meanwhile, it was also theoretically shown [10-13]
that magnetic properties of Q1D and Q2D superconductors can become
unique in very strong magnetic fields under conditions of the
so-called quantum $3D \rightarrow 2D$ dimensional crossovers,
where the typical "sizes" of electron trajectories are of the
order or less than the inter-plane distances.

The goal of our Letter is to introduce quantum $3D \rightarrow 1D
\rightarrow 2D$ dimensional crossover in a Q1D conductor and to
show that it can be responsible for the Fermi liquid - non Fermi
liquid crossovers (or phase transitions) in a tilted magnetic
field. We calculate inverse electron-electron scattering time and
demonstrate that it becomes almost 1D (i.e., of the order of the
characteristic electron energy, $1/\tau \sim \epsilon \sim T$) in
high magnetic fields, directed far from the main crystallographic
axes. In this case, Landau quasi-particles in Fermi liquid are not
well defined. Therefore, we can expect that Fermi liquid theory is
broken and some novel electronic states, including the possible
Luttinger liquid phase, appear. If magnetic field is directed
close to one of the main crystallographic axes, then, as we show
below, inverse electron-electron scattering time become 2D and,
thus, much less than the characteristic electron energy, $1 /\tau
\ll \epsilon \sim T$. In this case, we have to expect Fermi liquid
behavior of conducting electrons. It is important that in
(Per)$_2$Au(mnt)$_2$ layered Q1D conductor the above mentioned
Fermi liquid - non Fermi liquid crossovers (or transitions) are
expected to happen in feasibly high magnetic fields of the order
of $25 \ T$. We also discuss experimental results on investigation
of Lebed's magic angles in (Per)$_2$Au(mnt)$_2$ [14], where such
crossovers (or transitions) may have been already observed at $H
\simeq 30 \ T$.

Let us first demonstrate the suggested phenomenon, using
qualitative language. We consider a layered Q1D conductor with
electron spectrum, corresponding to the following two slightly
corrugated planes near $p_x = \pm p_F$:
\begin{equation}
\epsilon({\bf p})= \pm v_F(p_x \mp p_F) - 2 t_y \cos(p_y a_y) -
2t_z \cos (p_z a_z),
\end{equation}
where $p_F$ and $v_F$ are the Fermi momentum and Fermi velocity,
respectively; $p_F v_F \gg t_y \gg t_z$; $\hbar \equiv 1$.
Below, we study the case,
where magnetic field is perpendicular to the conducting chains and
makes angle $\alpha$ with the conducting planes,
\begin{equation}
{\bf H} = (0,\cos \alpha,\sin \alpha)H, \ {\bf A} = (0,-\sin
\alpha ,\cos \alpha ) Hx.
\end{equation}
To consider a quantum problem of the Q1D electrons (1) motion in
the magnetic field (2), we make use of the so-called Peierls
substitution method, formulated for a Q1D conductor in Ref.[4]. In
our particular case, this method allows to introduce magnetic
field by the following substitutions:
\begin{eqnarray}
p_x \mp p_F \rightarrow \mp i(d/dx), \ p_y a_y \rightarrow p_y a_y
- \omega_y(\alpha)/v_F,
\nonumber\\
p_z a_z \rightarrow p_z a_z + \omega_z(\alpha)/v_F ,
\end{eqnarray}
where
\begin{equation}
\omega_y(\alpha)=e v_F a_y H \sin \alpha /c, \ \
\omega_z(\alpha)=e v_F a_z H \cos \alpha /c \ .
\end{equation}
After these substitutions electron energy
(1) becomes the Hamiltonian operator and the corresponding
Schr\"{o}dinger-like equation for electron wave function in mixed
$(x; p_y,p_z)$ representation can be written as
\begin{eqnarray}
\biggl\{ \mp i v_F \frac{d}{dx} - 2t_y \cos
\biggl[p_ya_y-\frac{\omega_y(\alpha)}{v_F}x \biggl] - 2t_z \cos
\biggl[p_za_z
\nonumber\\
+\frac{\omega_z(\alpha)}{v_F}x \biggl] \biggl\}
\psi_{\epsilon}^{\pm}(x; p_y,p_z) = \delta \epsilon \
\psi_{\epsilon}^{\pm}(x; p_y,p_z),
\end{eqnarray}
where electron energy is counted from the Fermi level, $\delta
\epsilon = \epsilon - p_F v_F$. Note that Eq.(5) can be
analytically solved. As a result, we obtain:
\begin{eqnarray}
\psi_{\epsilon}^{\pm}(x; p_y,p_z) =  \exp \biggl( \frac{\pm i
\delta \epsilon x}{v_F} \bigg) \exp  \biggl\{ \mp i l_y(\alpha)
\sin \biggl[ p_y a_y
\nonumber\\
- \frac{\omega_y(\alpha)}{v_F} x \biggl] \biggl\} \exp \biggl\{
\pm i l_z (\alpha) \sin \biggl[ p_z a_z
 + \frac{\omega_z(\alpha)}{v_F} x \biggl]
 \biggl\},
\end{eqnarray}
where
\begin{equation}
l_y(\alpha) = \frac{2t_y}{\omega_y(\alpha)}, \ l_z(\alpha) =
\frac{2t_z}{\omega_z(\alpha)}.
\end{equation}
It is possible to show [3] that the parameters (7) are the "sizes"
of quasi-classical electron trajectories along $y$ and $z$ axes,
measured in terms of the corresponding lattice parameters, $a_y$
and $a_z$.

For further qualitative analysis it is
convenient to calculate the Fourier transformations of function
(6) for integer values of variables $y = n a_y$ and $z = m a_z$
(i.e., on the conducting chains):
\begin{eqnarray}
\psi_{\epsilon}^{\pm}(x,y=na_y,z=ma_z)=\int_0^{2\pi}\frac{d(p_ya_y)}{2\pi}
\exp(inp_ya_y)
\nonumber\\
\times \int_0^{2\pi}\frac{d(p_za_z)}{2\pi} \exp(imp_za_z)\
\psi_{\epsilon}^{\pm}(x,p_y,p_z).
\end{eqnarray}
After substitution of wave function (6) in Eq.(8) and
straightforward calculations it is possible to show that
\begin{eqnarray}
\psi_{\epsilon}^{\pm}(x,y=na_y,z=ma_z)& = \exp \biggl\{ \frac{\pm
i[\delta \epsilon \pm n \omega_y(\alpha) \mp m
\omega_z(\alpha)]x}{v_F} \biggl\}
\nonumber\\
&\times J_n[\pm l_y(\alpha)] \ J_m[\mp l_z(\alpha)],
\end{eqnarray}
where we make use of the following property of the Bessel
functions [15]:
\begin{equation}
J_n(z) = \int_{- \pi}^{\pi} \frac{d \phi}{2 \pi} \ \exp(i n \phi)
\ \exp[- i z \sin (\phi)] \ .
\end{equation}
We note that wave function in a real space (9) show the amplitudes
for an electron to occupy the conducting chains with the
coordinates $y=na_y$ and $z=ma_z$ in a Q1D conductor in case, where
electron wave function is centered at $y = z = 0$. In particular,
from Eq.(9), it follows that the total probability to occupy all
possible chains at arbitrary magnetic field is
\begin{equation}
P=\sum_{n=-\infty}^{+\infty} \sum_{m=-\infty}^{+\infty} \
J^2_n[\pm l_y(\alpha)]\ J^2_m[\mp l_z(\alpha)]= 1 ,
\end{equation}
as it has to be, where we use that $\sum_{n=-\infty}^{+\infty}
J_n^2(z)=1$ for arbitrary value of the argument $z$ [15].

Note that wave functions in a real space (9) are although one
dimensional but in general occupy many conducting chains.
Nevertheless, when the parameters (7) become smaller than 1 in
high magnetic fields,
\begin{equation}
H \geq H^* = \max \biggl\{ \frac{2t_y c}{ev_Fa_y \sin \alpha} \ ,
\frac{2t_z c}{ev_Fa_z \cos \alpha} \biggl\} \ ,
\end{equation}
electron wave functions (9) become localized on the conducting
chain with $y=z=0$. This fact is directly seen from the following
properties of the Bessel functions [15]:
\begin{equation}
\lim_{z \rightarrow 0} \ J_0(z) \rightarrow 1 ; \
\lim_{z \rightarrow 0} \ J_n(z) \rightarrow 0, \ n \neq 0.
\end{equation}
The above mentioned localization of electrons means that high
enough magnetic fields fully "one-dimensionalize" Q1D electron
spectrum (1). Therefore, we expect that, at high magnetic fields,
Q1D electrons start to exhibit non Fermi liquid properties, since
Fermi liquid is known not to exist in a pure $1D$ case. It is
important that this can happen only in the case, where direction
of a magnetic field is far from the crystallographic axes at
$\alpha = 0^o$ and $\alpha = 90^o$. Indeed, if direction of a
magnetic field is close to one of these axes, the "sizes" (7) of
electron wave function (9) become large and, thus, Fermi liquid
properties have to be restored. Therefore, we expect Fermi liquid
- non Fermi liquid angular crossovers (or phase transitions) in a
tilted high enough magnetic field. Let us estimate the value of
the critical magnetic field (12) for the Q1D conductor
(Per)$_2$Au(mnt)$_2$, using the following values for the
parameters of its electron spectrum [14]: $v_F = 1.7 \times 10^7
cm/s$, $t_y = 20 \ K$, $t_z \leq t_y$, $a_y = 16.6 \ \AA$, and
$a_z=30 \ \AA$. In this case, from Eq.(12), we estimate that $H^*
\simeq 25 \ T$ at $\alpha = 45^0$ - the value, which is available
as a steady magnetic field.

Below, we calculate inverse electron-electron scattering time and
directly demonstrate that the major Landau criterion [16,17] for
Fermi liquid behavior is broken at high enough magnetic fields
(12). We recall that this criterion says that Landau
quasi-particles have to be well defined in Fermi liquid. In
particular, this means that electron-electron scattering time has
to be much less than the typical electron energy, $1/\tau \ll
\epsilon \sim T$. For further calculations, it is important that,
in a magnetic field (2), only electron momenta, perpendicular to
the conducting chains, $p_y$ and $p_z$, are conserved. This means
that momentum conservation law can be written in the collision
integral for Fermi particles [16,17] only for the above mentioned
directions. On the other hand, the total electron energy is
conserved in a magnetic field. In order, to calculate inverse
electron-electron scattering time, averaged over electron energy,
$\epsilon$, and perpendicular components of momentum, $p_y$ and
$p_z$, we need to consider the following expression, extended
electron-electron collision integral to the case of
non-conservation of momentum along conducting axis ${\bf x}$:

\begin{eqnarray}
&\frac{1}{\tau} = \int d \epsilon_1 \int d \epsilon_2 \int d
\epsilon_3 \int d \epsilon_4 \ \delta(\epsilon_1 + \epsilon_2
-\epsilon_3 - \epsilon_4) \nonumber\\
&\times
n(\epsilon_1)n(\epsilon_2)[1-n(\epsilon_3)][1-n(\epsilon_4)]
\nonumber\\
&\times \int dp^1_y \int dp^2_y \int d q_y \int dp^1_z \int dp^2_z
\int d
q_z \nonumber\\
&W(\epsilon_1,p^1_y,p^1_z;\epsilon_2,p^2_y,p^2_z; \nonumber\\
&\epsilon_3,p^1_y+q_y,p^1_z+q_z; \epsilon_4,p^2_y-q_y,p_z^2-q_z).
\end{eqnarray}
To find electron-electron scattering probability, $W(...)$ in
Eq.(14), in a magnetic field (2), we make use of known electron
wave functions (8). It is possible to show that the scattering
probability, corresponding to electron-electron scattering
amplitude, shown in Fig.1, is
\begin{eqnarray}
&W(\epsilon_1,p^1_y,p^1_z;\epsilon_2,p^2_y,p^2_z;\epsilon_3,p^1_y+q_y,p^1_z+q_z;
\epsilon_4,p^2_y-q_y,p_z^2-q_z) \nonumber\\
&= U \int dx \exp[i(\epsilon_1 - \epsilon_2 + \epsilon_3 -
\epsilon_4)x/v_F] \nonumber\\
&\times \exp \biggl\{ 8i l_y(\alpha) \sin
\bigl[\frac{\omega_y(\alpha)x}{2v_F}\bigl] \sin
\bigl[\frac{p^1_y+p^2_y}{2}\bigl] \cos
\bigl[\frac{p^1_y-p^2_y}{2}\bigl] \sin \bigl[\frac{q_y}{2}\bigl]
\biggl\} \nonumber\\
&\times \exp \biggl\{ 8i l_z(\alpha) \sin
\bigl[\frac{\omega_z(\alpha)x}{2v_F}\bigl] \sin
\bigl[\frac{p^1_z+p^2_z}{2}\bigl] \cos
\bigl[\frac{p^1_z-p^2_z}{2}\bigl] \sin \bigl[\frac{q_z}{2}\bigl]
\biggl\}.
\end{eqnarray}
We point out that we use approximation, where electron-electron
interaction, $U$, is independent on electron momenta in the
absence of a magnetic field, which corresponds to
electron-electron interaction term proportional to $\delta^3({\bf
r_1}-{\bf r_2})$ in a real space. In this case, all possible
amplitudes of electron-electron scattering give the same
probability (15).

After lengthly but straightforward calculations, we obtain from
Eq.(15):

\begin{eqnarray}
&\frac{1}{\tau} = 2 g^2 T \int_0^{\infty} \ \biggl( \frac{2 \pi T
dx}{v_F} \biggl) \biggl[ \frac{(\frac{2\pi T x}{v_F}) \cosh(\frac{2
\pi T x}{v_F}) - \sinh( \frac{2 \pi T x}{v_F})}{\sinh^3(\frac{ 2
\pi T x}{v_F})} \biggl]
\nonumber\\
&\times \biggl< J_0^2\biggl\{4 l_y(\alpha) \sin \biggl[
\frac{\omega_y(\alpha) x}{2v_F} \biggl] \cos(\phi_1) \biggl\}
\biggl>_{\phi_1}
\nonumber\\
&\times \biggl< J_0^2\biggl\{4 l_z(\alpha) \sin \biggl[
\frac{\omega_z(\alpha) x}{2v_F} \biggl] \cos(\phi_2) \biggl\}
\biggl>_{\phi_2} ,
\end{eqnarray}
\begin{figure}[h]
\centering
\includegraphics[width=85mm, height=55mm]{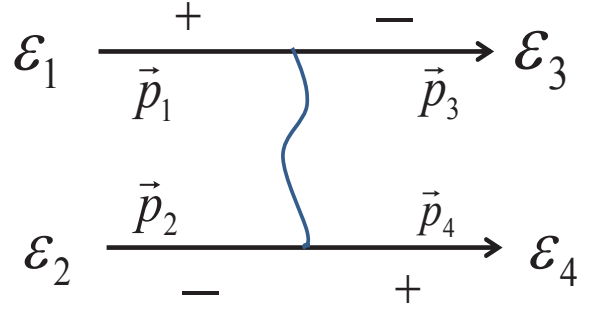}
\caption{One possible amplitude of electron-electron scattering,
where the first electron is scattered from right sheet of the Q1D
Fermi surface (1) to left sheet, whereas the second electron is
scattered from left sheet to right sheet.}
\end{figure}
where $<...>_{\phi}$ denotes averaging over variable $\phi$, $g$
stands for dimensionless electron-electron interaction constant.
Note that the inverse electron-electron scattering time (16) is
normalized in such a way that $1/\tau = g^2 T$ in a pure 1D case.
We point out that in the derivation of Eq.(16), we make use of
Eq.(10) as well as the following equations [15]:
\begin{eqnarray}
&\int \int \int \int d \epsilon_1 d \epsilon_2 d \epsilon_3 d
\epsilon_4
\exp[i(\epsilon_1-\epsilon_2+\epsilon_3-\epsilon_4)x/v_F]
\nonumber\\
&\times \delta(\epsilon_1+\epsilon_2-\epsilon_3-\epsilon_4)
n(\epsilon_1) n(\epsilon_2) [1-n(\epsilon_3)] [1-n(\epsilon_4)]
\nonumber\\
&=2 \pi^2 T^3 \biggl[ \frac{(\frac{2 \pi T x}{v_F}) \cosh (\frac{2
\pi T x }{v_x})-\sinh(\frac{2 \pi T x}{v_F})}{\sinh^3(\frac{2 \pi
T x}{v_F})}\biggl]
\end{eqnarray}
and
\begin{equation}
\int_0^{2 \pi} \frac{d \phi}{2 \pi} J_0(2z \sin \phi)=J^2_0(z).
\end{equation}

Let us analyze Eq.(16) at high magnetic fields. First, we consider
the case, where magnetic field is directed far from the
crystallographic axes at $\alpha = 0^0$ and $\alpha =90^0$. If
magnetic field satisfies Eq.(12), as it directly follows from
Eq.(16), both Bessel functions are of the order of 1 and inverse
electron-electron scattering time is of the order of
\begin{equation}
\frac{1}{\tau} \sim g^2 T \sim T.
\end{equation}
In other words, at high magnetic fields (12), inverse
electron-electron scattering time is completely
"one-dimensionalized" (i.e., becomes of the order of the
characteristic electron energy, $\frac{1}{\tau} \sim \epsilon \sim
T$). According to Landau [16,17], in this case the notion of
quasi-particles in Fermi liquid loses its meaning. Therefore,
under these conditions, we expect non Fermi liquid behavior of the
Q1D electron gas (1). Now, let us consider inverse
electron-electron scattering time (16) in the case, where magnetic
field is applied along ${\bf y}$ axis, which corresponds to
$\alpha = 0^0$. In this case, integral (16) can be estimated as
\begin{eqnarray}
&\frac{1}{\tau}(0^0) = 2 g^2 T \int_0^{\infty} \ \biggl( \frac{2
\pi T dx}{v_F} \biggl) \biggl[ \frac{\frac{2\pi T x}{v_F}
\cosh(\frac{2 \pi T x}{v_F}) - \sinh( \frac{2 \pi T
x}{v_F})}{\sinh^3(\frac{ 2 \pi T x}{v_F})} \biggl]
\nonumber\\
&\times \lim_{\alpha \rightarrow 0} \biggl< J_0^2\biggl\{4
l_y(\alpha) \sin \biggl[ \frac{\omega_y(\alpha) x}{2v_F} \biggl]
\cos(\phi_1) \biggl\} \biggl>_{\phi_1} .
\end{eqnarray}
Let us consider the case of low enough temperatures, where
\begin{equation}
T \ll t_y \simeq \omega_y(\alpha = 90^0).
\end{equation}
In this case for small enough angles,
\begin{equation}
\sin \alpha \ll T / \omega_y(\alpha = 90^0),
\end{equation}
integral (20) can be simplified as
\begin{eqnarray}
&\frac{1}{\tau}(0^0) = 2 g^2 T \int_0^{\infty} \ dz \frac{z
\cosh(z) - \sinh(z)}{\sinh^3(z)}
\nonumber\\
&\times \biggl< J_0^2\biggl( \frac{2 t_y z}{\pi T} \cos \phi \biggl)
\biggl>_{\phi} .
\end{eqnarray}
Note that the integral (23) can be analytically calculated with
the so-called logarithmic accuracy:
\begin{equation}
\frac{1}{\tau}(0^0) \simeq  \frac{ g^2 T^2}{2 \pi t_y} \ln^2 \biggl(
\frac{t_y}{T} \biggl) \ll T .
\end{equation}
As it follows from Eq.(24), for small enough angles (22) inverse
electron-electron scattering time becomes smaller than the
electron characteristic energy, $1/\tau \ll \epsilon \sim T$, and
the concept of quasi-particles in Fermi liquid restores [16,17].
Therefore, we expect restoration of Fermi liquid behavior for
$\alpha \simeq 0^0$. In Fig.2, the results of careful numerical
calculations of Eq.(16) are presented, which confirm the above
mentioned analytical analysis. To obtain inverse electron-electron
relaxation time for magnetic field, directed close to ${\bf z}$
axis ($\alpha = 90^0$), we need to do the following substitutions
$t_y \rightarrow t_z$ and $\omega_y(\alpha) \rightarrow
\omega_z(\alpha)$ in Eqs.(22),(24). As a result, we obtain
\begin{equation}
\frac{1}{\tau}(90^0) \simeq  \frac{ g^2 T^2}{2 \pi t_z} \ln^2 \biggl(
\frac{t_z}{T} \biggl) \ll T
\end{equation}
for
\begin{equation}
\cos \alpha \ll T / \omega_z(\alpha = 0^0),
\end{equation}
and, thus, Fermi liquid behavior is expected to restore also at
angles close to $90^0$. [We note that there are some mathematical
similarities between microscopic problem, considered in this
Letter, and semi-phenomenological calculations of conductivity of
a Q1D metal in a magnetic field [18]. Nevertheless, the physical
conclusions of our Letter and Ref.[18] are quiet different.]
\begin{figure}[h]
\centering
\includegraphics[width=85mm, height=55mm]{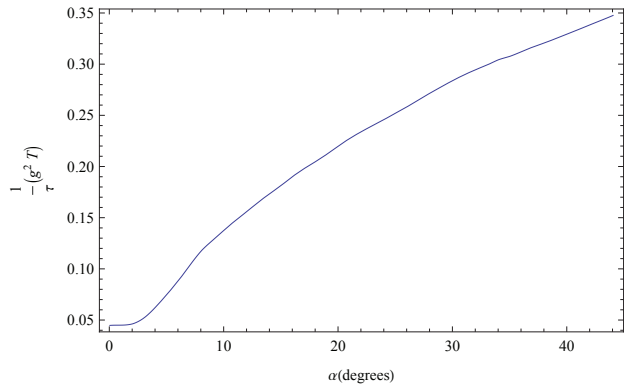}
\caption{Inverse electron-electron scattering time, calculated by
means of Eq.(16) and expressed in term of $g^2 T$, is shown as a
function of angle $\alpha$. The calculations are done for the
parameters $l_y(0)=1$, $t_z/t_y = 0.2$, $a_y = a_z/2$, $4 \pi
T/\omega_y(0) = 0.2$, which correspond to $H \simeq 25 \ T$.}
\end{figure}

In conclusion, we discuss possible experimental applications of
the suggested above Fermi liquid - non Fermi liquid angular
crossovers (or phase transitions) in a Q1D conductor in high
magnetic fields. The most natural way is to perform the
corresponding experiments in the Q1D conductor
(Per)$_2$Au(mnt)$_2$ under pressure, where charge-density-wave
state is destroyed and metallic Fermi liquid phase is a ground
state at $H=0$ [14,19]. In addition to resistive experiments
[14,19], we suggest also torque measurements in high magnetic
fields, $H \simeq 25 \ T$, perpendicular to the conducting chains,
since the angular Fermi liquid - non Fermi liquid crossovers have
to have also thermodynamic consequence [20]. In this context, we
note that, as shown by Yakovenko [21], Lebed's Magic angle effects
have to exist already in moderate magnetic fields in Fermi liquid
phase of a Q1D conductor for different thermodynamic properties
such as torque, specific heat, and magnetic moment. As to the
resistive measurements of Lebed's Magic angle phenomenon [14], it
seems that one feature of the above mentioned crossovers has been
already observed in Ref.[14] - non-metallic temperature dependence
of resistance for high magnetic fields, directed far from the main
crystallographic axes. It is important that this non-metallic
behavior cannot be a consequence of Fermi liquid
magnetoresistance, since, for experimental current along the
conducting axis, ${\bf I} \parallel {\bf b}$, Fermi liquid
magnetoresistance is expected to be zero. On the other hand, it is
known that it is not easy to measure conductivity in a Q1D
conductor exactly along its conducting axis [3], therefore, more
experimental works are needed.

We stress that the effects, suggested in the Letter, are rather
general and have to be observed in other materials containing Q1D
parts of the Fermi surfaces, such as (TMTSF)$_2$X salts and some
BEDT-based materials. Nevertheless, the required magnetic field
for Fermi liquid - non Fermi liquid crossovers in the above
mentioned conductors are estimated as $H^* \simeq 250 \ T$, which
is order of magnitude higher than the value for the Q1D
(Per)$_2$Au(mnt)$_2$ conductor.

We are thankful to N.N. Bagmet for useful discussions. This work
was supported by the NSF under Grant No DMR-1104512.

$^*$Also at: L.D. Landau Institute for Theoretical Physics, RAS,
2 Kosygina Street, Moscow 117334, Russia.

\end{document}